\documentclass[AMA,Times1COL]{WileyNJDv5} 

\usepackage{amsmath}
\usepackage{graphicx,psfrag,epsf}
\usepackage{enumerate}
\usepackage{url}
\usepackage{algpseudocode}
\usepackage{algorithm}
\usepackage{amsfonts}

\articletype{Article Type}%

\received{Date Month Year}
\revised{Date Month Year}
\accepted{Date Month Year}
\journal{Statistics in Medicine}
\volume{00}
\copyyear{2025}
\startpage{1}

\begin{document}

\title{Efficient Computation of High-Dimensional Penalized Piecewise Constant Hazard Random Effects Models}

\author[1]{Hillary M. Heiling}

\author[1,2]{Naim U. Rashid}

\author[1]{Quefeng Li}

\author[2]{Xianlu L. Peng}

\author[2,3,4]{Jen Jen Yeh}

\author[1]{Joseph G. Ibrahim}

\authormark{HEILING \textsc{et al.}}
\titlemark{EFFICIENT COMPUTATION OF HIGH-DIMENSIONAL PENALIZED PIECEWISE CONSTANT HAZARD RANDOM EFFECTS MODELS}


\address[1]{\orgdiv{Department of Biostatistics}, \orgname{University of North Carolina at Chapel Hill}, \orgaddress{\state{North Carolina}, \country{United States of America}}}

\address[2]{\orgdiv{Lineberger Comprehensive Cancer Center}, \orgname{University of North Carolina at Chapel Hill}, \orgaddress{\state{North Carolina}, \country{United States of America}}}

\address[3]{\orgdiv{Department of Surgery}, \orgname{University of North Carolina at Chapel Hill}, \orgaddress{\state{North Carolina}, \country{United States of America}}}

\address[4]{\orgdiv{Department of Pharmacology}, \orgname{University of North Carolina at Chapel Hill}, \orgaddress{\state{North Carolina}, \country{United States of America}}}

\corres{Corresponding author Hillary Heiling. \email{hillary\_heiling@dfci.harvard.edu}}



\abstract[Abstract]{Identifying and characterizing relationships between treatments, exposures, or other covariates and time-to-event outcomes has great significance in a wide range of biomedical settings. In research areas such as multi-center clinical trials, recurrent events, and genetic studies, proportional hazard mixed effects models (PHMMs) are used to account for correlations observed in clusters within the data. In high dimensions, proper specification of the fixed and random effects within PHMMs is difficult and computationally complex. In this paper, we approximate the proportional hazards mixed effects model with a piecewise constant hazard mixed effects survival model.  We estimate the model parameters using a modified Monte Carlo Expectation Conditional Minimization algorithm, allowing us to perform variable selection on both the fixed and random effects simultaneously. We also incorporate a factor model decomposition of the random effects in order to more easily scale the variable selection method to larger dimensions. We demonstrate the utility of our method using simulations, and we apply our method to a multi-study pancreatic ductal adenocarcinoma gene expression dataset to select features important for survival.}


\jnlcitation{\cname{%
\author{Taylor M.},
\author{Lauritzen P},
\author{Erath C}, and
\author{Mittal R}}.
\ctitle{On simplifying ‘incremental remap’-based transport schemes.} \cjournal{\it J Comput Phys.} \cvol{2021;00(00):1--18}.}

\maketitle

\renewcommand\thefootnote{}
\footnotetext{\textbf{Abbreviations:} GLMM, generalized linear mixed model; MCECM, Monte Carlo expectation conditional minimization; PHMM, proportional hazards mixed model}

\renewcommand\thefootnote{\fnsymbol{footnote}}
\setcounter{footnote}{1}

\section{Introduction}
\label{sec:proj3_intro}

Modeling survival outcomes has great clinical significance in medical and public health research. In particular, the Cox proportional hazards model has been widely utilized in order to characterize the relationship between treatments, exposures, or other covariates and patient time-to-event outcomes. However, modern biomedical datasets are increasingly high dimensional, and groups of samples within the data can exhibit complex correlations. For example, when studying survival outcomes with respect to multi-center clinical trials, recurrent events, and genetic studies, proportional hazards mixed effects models are used to account for correlations among groups within the data and model the heterogeneity of treatment and predictor effects across groups.\citep{vaida2000proportional,ripatti2000estimation} These proportional hazards mixed effects models are traditionally referred to as frailty models when the model contains a single random effect applied to the baseline hazard. 

In high dimensional settings, in which the covariate effects are generally assumed to be sparse, it is often unknown \textit{a priori} which covariates should be specified as fixed or random in the model. Variable selection methods such as LASSO and SCAD exist for high dimensional proportional hazards models or frailty models,\citep{tibshirani1997lasso,bradic2011regularization,glmnet_coxnet2011,fan2002variable} but they do not allow for the selection of random effects. Several mixed effects model selection methods that rely on the specification of candidate models have been proposed, including likelihood ratios, profile Akaike information criterion (AIC),\citep{xu2009using} and conditional AIC.\citep{donohue2011conditional} However, specifying all $2^p$ possible candidate models in high dimensions is impractical. Lee et al. \cite{lee2014bayesian} developed a stochastic search variable selection (SSVS) method that selects both fixed and random effects in proportional hazards mixed effects models in a Bayesian framework, but their method is only computationally feasible for small or moderate dimensions.

Rashid et al. \cite{rashid2020} developed a method that simultaneously selects both fixed and random effects in high dimensional generalized linear mixed models (GLMMs), which has since been developed into an R package available on CRAN, called \textbf{glmmPen}.\citep{glmmPen_pkg,heiling2023glmmpen} This method broadened the feasible dimensionality of performing variable selection in GLMMs to greater dimensions than previously existing methods. This method was extended by Heiling et al.,\cite{heiling2024efficient} who proposed a new formulation of the GLMM using a factor model decomposition of the random effects. As a result of this new formulation, they were able to improve the scalability of their method and perform variable selection within GLMMs in cases with much larger dimensions. However, these methods do not apply to survival data.


In this paper, we propose a method that simultaneously selects fixed and random effects within clustered survival data. In order to extend the methods of Rashid et al. \cite{rashid2020} and Heiling et al. \cite{heiling2024efficient} to survival data, we first consider an approximation of the Cox proportional hazards model using a piecewise constant hazard model. Piecewise constant hazard models, sometimes referred to as piecewise exponential models, are parametric survival models that divide the follow-up time of the data into intervals, where the hazard function is assumed to be constant in each interval; this piecewise constant hazard survival model can be fit using a log-linear model which incorporates the duration of exposure within each interval.\citep{allison2010survival,holford1980analysis,laird1981covariance,friedman1982piecewise}
Piecewise constant hazard models can be extended to include random effects using piecewise constant hazard mixed effects models.\citep{ibrahim2001bayesian,austin2017tutorial,gray1994bayesian} In our piecewise constant hazard mixed effects model, we utilize the factor model decomposition of the random effects proposed in Heiling et al. \cite{heiling2024efficient}, allowing us to scale our method to cases with hundreds of predictors. We label our method as  \textbf{phmmPen\_FA}, which reflects our goal of estimating penalized proportional hazards mixed effects models using factor analysis on the random effects.  

We demonstrate the application of our method by applying our method to a case study that we present in Section \ref{sec:proj3_PDAC}. The development of this method was motivated by a case study dataset that contains gene expression data from pancreatic ductal adenocardinoma patients across seven separate studies. We aim to select important features that predict subjects' survival by identifying features that increase or decrease the rate of the occurrence of death (i.e. identify features with non-zero fixed effect hazard ratios) as well as identify features that have varied effects on subjects' survival across the groups (i.e. identify non-zero random effects). Due to the large number of features that we consider, it is difficult to have \textit{a priori} knowledge of which features have non-zero fixed or random effects. Therefore, we will use the \textbf{phmmPen\_FA} method to fit a penalized piecewise constant hazard mixed effects survival model to select important fixed and random effects.

The remainder of this paper is organized as follows. 
Section \ref{sec:proj3_methods} reviews the statistical models and algorithm used to estimate piecewise constant hazard mixed effects models. In section \ref{sec:proj3_sims}, simulations are conducted to assess the performance of our method. Section \ref{sec:proj3_PDAC} describes the motivating case study for the prediction of survival in pancreatic ductal adenocarinoma cancer using gene expression data from multiple studies, and provides results from the application of our new method to the case study. We close the article with some discussion in Section \ref{sec:proj3_discuss}. 

Software in the form of R code for the \textbf{phmmPen\_FA} procedure is available through the \textbf{glmmPen} package in CRAN \url{https://cran.r-project.org/package=glmmPen} and the GitHub repository \url{https://github.com/hheiling/glmmPen}. The \textbf{phmmPen\_FA} procedure is implemented through the \texttt{phmmPen\_FA()} function within this \textbf{glmmPen} R package. 

\section{Methods}
\label{sec:proj3_methods}

\subsection{Model formulation}
\label{sec:proj3_models}


In this section, we review the notation and model formulation of our approach. We begin by introducing the proportional hazards mixed effects model using a generic linear predictor with both fixed and random effects. We then discuss how we approximate this model with a piecewise constant hazard mixed effects model, which is similar to how fixed-effects only Cox proportional hazards models are approximated with piecewise constant hazard models.\citep{ibrahim1999bayesian,austin2017tutorial,allison2010survival,gray1994bayesian} Next, we explain how we represent this model using a log-linear mixed effects model.\citep{austin2017tutorial,holford1980analysis,laird1981covariance} Once we have set up the general model scheme, we provide details about the exact form of our assumed linear predictor, where we incorporate the factor model decomposition of the random effects proposed in Heiling et al.\cite{heiling2024efficient} Lastly, we introduce penalties into the model and discuss how the variable selection goals of our algorithm relate to the introduced notation. Details about the MCECM algorithm we use to fit this model are discussed in Section \ref{sec:proj3_mcecm}.

We consider the case where we want to analyze data from \(K\) independent groups of subjects. For each group \(k = 1,...,K\), there are \(n_k\) subjects for a total sample size
of \(N = \sum_{k=1}^K n_k\). For group $k$, let
\(\boldsymbol y_k = (y_{k1},...,y_{kn_k})^T\) be the vector of \(n_k\)
observed times, where $y_{ki} = \min(T_{ki},C_{ki})$, $T_{ki}$ represents the event time, and $C_{ki}$ represent censoring time; let $\boldsymbol \delta_k = (\delta_{k1},...,\delta_{kn_k})$ where $\delta_{ki} = I(T_{ki} < C_{ki})$ represents the indicator that a subject's event time was observed; and let \(\boldsymbol x_{ki} = (x_{ki,1},...,x_{ki,p})^T\) be the
\(p\)-dimensional vector of predictors, and \(\boldsymbol X_k = (\boldsymbol x_{k1}, ..., \boldsymbol x_{kn_k})^T\). We standardize the fixed effects covariates matrix
\(\boldsymbol X = (\boldsymbol X_1^T,...,\boldsymbol X_K^T)^T\) such that \\ \(\sum_{k=1}^K \sum_{i=1}^{n_{k}} x_{ki,j} = 0\) and
\(N^{-1} \sum_{k=1}^K \sum_{i=1}^{n_{k}} x_{ki,j}^2 = 1\) for
\(j = 1,...,p\).

We would like to estimate the proportional hazards mixed effects model
\begin{equation}
    h(t|\eta_{ki}) = h_0(t) \exp(\eta_{ki}),
    \label{eqn:chp5_coxph}
\end{equation}
where $h(t|\eta_{ki})$ is the individual hazard of subject $i$ in group $k$ at time $t$, $h_0(t)$ represents the baseline hazard at time $t$, and $\eta_{ki}$ represents the linear predictor containing the fixed effects log hazard ratio coefficients, the group-specific random effects, and the subject's individual covariates. The exact form of the linear predictor $\eta_{ki}$ assumed in our model is described later in this section.

In cases where we are modeling survival data with only fixed effects in the model (i.e. no random effects) and we are not using Bayesian techniques to estimate these fixed effects, we can ignore the baseline hazard function $h_0(t)$ during the estimation of the fixed effects coefficients in the linear predictor. However, when survival models include random effects or involve estimation using Bayesian techniques,
it is necessary to fully specify and model the baseline hazard function. In this paper, we approximate the baseline hazard using a piecewise constant function \citep{allison2010survival} such that we approximate the model (\ref{eqn:chp5_coxph}) using the piecewise constant hazard mixed effects model.\citep{austin2017tutorial} We use this approximation because modeling the baseline hazard as a piecewise constant function in this setting allows for relatively convenient computation.
We first partition the time of the study into $J$ intervals, where we assume that the baseline hazard within a particular time interval is constant. Please see Section \ref{sec:proj3_sims} for a discussion on choosing the number of time intervals $J$. Let us define the cut points $0 = \tau_0 < \tau_1 < ... < \tau_J = \infty$, and let $h_j$ be the constant baseline hazard within interval $j$, $[\tau_{j-1},\tau_j)$. We then approximate (\ref{eqn:chp5_coxph}) using the model  
\begin{equation}
    h_{kij} = h_j \exp(\eta_{ki}),
    \label{eqn:chp5_piecewise_haz_model}
\end{equation}
where $h_j$ is the baseline hazard for interval $j$ and $h_{kij}$ is the constant hazard corresponding to subject $i$ in group $k$ within interval $j$.

The observed data for each subject includes their observed time $y_{ki}$ and their event indicator $\delta_{ki}$. We extend these to define analogous measures for each interval, where $t^*_{kij} = \max[\min(y_{ki},
\tau_j) - \tau_{j-1}, 0]$ is the amount of time subject $i$ in group $k$ survived within interval $j$, and $d_{kij} = I(\tau_{j-1} \leq y_{ki} < \tau_j, \delta_{ki} = 1)$ is the indicator of whether the subject died during interval $j$. To better clarify $t^*_{kij}$, this term has three possible values, determined by the relative value of their observed time $y_{ki}$ to the interval cut points:
\begin{equation*}
    t^*_{kij} = 
    \begin{cases}
        \tau_j - \tau_{j-1}, & y_{ki} > \tau_j; \\
        y_{ki} - \tau_{j-i}, & \tau_{j-1} < y_{ki} \leq \tau_j; \\
        0, & y_{ki} \leq \tau_{j-1}.
    \end{cases}
\end{equation*}
We can then treat the death indicators $d_{kij}$ as if they are independent Poisson observations with means $\mu_{kij} = t^*_{kij} h_{kij}$.
Consequently, 
we estimate the fixed and random effect parameters of our model by fitting the data using the log-linear model 
\begin{equation}
    \log \mu_{kij} = \log t^*_{kij} + \psi_j + \eta_{ki},
    \label{eqn:chp5_pois_model}
\end{equation}
where $\psi_j = \log(h_j)$ is the logarithm of the constant hazard within interval $j$ and $\log(t^*_{kij})$, which represents the log of the time a subject survived within interval $j$, is treated as an offset to the model. We treat the $d_{kij}$ values as the outcome (i.e. response) values of model (\ref{eqn:chp5_pois_model}) for each subject $i$ in group $k$ for each interval $j$.

Let us define $\boldsymbol{d}_k = (d_{k11},...,d_{k1J},...,d_{kn_k1},...,d_{kn_kJ})^T$ as the vector of death indicator values for all subjects in group $k$ and all $J$ time intervals. 
Then, the piecewise constant hazard likelihood is defined as
\begin{align}
  f(\boldsymbol d_k | \boldsymbol X_k, \boldsymbol \alpha_k; \theta) 
    &= \prod_{i=1}^{n_k} \prod_{j=1}^J \left [I(t^*_{kij} > 0) \mu_{kij} \right ]^{d_{kij}} \exp[- I(t^*_{kij} > 0) \mu_{kij}],
    \label{eqn:chp5_PE_likelihood}
\end{align}
where $\mu_{kij}$ is defined in (\ref{eqn:chp5_pois_model}), and $I(t^*_{kij} > 0)=1$ indicates that a subject $i$ in group $k$ survived at least part way through interval $j$, 0 if the subject died or was censored before interval $j$.

Now we may defined the form of the linear predictor term $\eta_{ki}$ used within this model. We start by introducing the linear predictor term used in the traditional generalized linear mixed model \citep{chen2003,BICq2011,rashid2020}
\begin{equation}
  \eta_{ki} = \boldsymbol x_{ki}^T \boldsymbol \beta + \boldsymbol z_{ki}^T \boldsymbol \gamma_k = \boldsymbol x_{ki}^T \boldsymbol \beta + \boldsymbol z_{ki}^T \boldsymbol \Gamma \boldsymbol \epsilon_k,
  \label{eqn:chp5_linpredgam}
\end{equation}
where \(\boldsymbol \beta = (\beta_1,...,\beta_{p})^T\) is a \(p\)-dimensional vector for the fixed effects coefficients ($\boldsymbol \beta$ represents the log hazard ratio values for each predictor and excludes an intercept, just as in a typical Cox proportional hazards model), $\boldsymbol \Gamma$ is the Cholesky decomposition of the random effects covariance matrix $\boldsymbol \Sigma$ such that $\boldsymbol \Gamma \boldsymbol \Gamma^T = \boldsymbol \Sigma$,
$\boldsymbol \gamma_k = \boldsymbol \Gamma \boldsymbol \epsilon_k$ is a \(q\)-dimensional vector of unobservable random effects (including the random intercept) for group $k$ where $\boldsymbol \epsilon_k \sim N_q(0,\boldsymbol I)$, and \(\boldsymbol z_{ki}\) is a \(q\)-dimensional vector that includes an intercept term and a subset of \(\boldsymbol x_{ki}\).

We reformulate the linear predictor of (\ref{eqn:chp5_linpredgam}) using the technique described in Heiling et al.,\cite{heiling2024efficient} where we decompose the random effects $\boldsymbol \gamma_k$ into a factor model with $r$ latent common factors and we assume $r \ll q$. As a result, we assume $\boldsymbol \gamma_k = \boldsymbol B \boldsymbol \alpha_k$, where $\boldsymbol B$ is the $q \times r$ loading matrix and $\boldsymbol \alpha_k$ represents the $r$ latent common factors. We assume the latent factors $\boldsymbol \alpha_k$ are uncorrelated and follow a $N_r(\boldsymbol 0, \boldsymbol I)$ distribution. We re-write the linear predictor as
\begin{equation}
  \eta_{ki} = \boldsymbol x_{ki}^T \boldsymbol \beta + \boldsymbol z_{ki}^T \boldsymbol B \boldsymbol \alpha_k.
  \label{eqn:chp5_linpredB}
\end{equation}
In the representation of (\ref{eqn:chp5_linpredB}), the random component of the linear predictor has variance Var(\(\boldsymbol B \boldsymbol \alpha_k\)) = \(\boldsymbol{B B}^T\) = $\boldsymbol \Sigma$, which is low rank. By using this representation, we reduce the dimension of the latent space from $q$ to $r$. As a result, this reduces the dimension of the integral in the likelihood, which reduces the computational complexity of the E-step in the EM algorithm described in Section \ref{sec:proj3_mcecm}. Consequently, this factor decomposition reduces the computational time of the algorithm and enables our method to scale to hundreds of predictors.\citep{heiling2024efficient}

In order to estimate \(\boldsymbol B\), let $\boldsymbol b_t \in \mathbb{R}^r$ be the $t$-th row of $\boldsymbol B$ and $\boldsymbol b = (\boldsymbol b_1^T,...,\boldsymbol b_q^T)^T$. 
We then reparameterize the linear predictor as
\begin{equation}
  \eta_{ki} = \boldsymbol x_{ki}^T \boldsymbol\beta + \boldsymbol z_{ki}^T \boldsymbol B \boldsymbol\alpha_k = \left (\boldsymbol x_{ki}^T \hspace{10 pt} (\boldsymbol\alpha_k \otimes \boldsymbol z_{ki})^T \boldsymbol J^\diamond \right)
  \left (
  \begin{matrix}
    \boldsymbol\beta \\ \boldsymbol b 
  \end{matrix}
  \right )
\end{equation} in a manner similar to Chen and Dunson \cite{chen2003} and Ibrahim et al.,\cite{BICq2011} where \(\boldsymbol J^\diamond\) is a matrix that transforms $\boldsymbol b$ to vec($\boldsymbol B$) such that $\text{vec}(\boldsymbol B) = \boldsymbol J^\diamond \boldsymbol b$ and \(\boldsymbol J^\diamond\) is of
dimension $(qr)\times(qr)$. The vector of parameters
\(\boldsymbol \theta = (\boldsymbol \beta^T, \boldsymbol b^T, \boldsymbol \psi^T)^T\)
are the main parameters of interest. 

We denote the true value of
\(\boldsymbol \theta\) as
\(\boldsymbol \theta^{*} = (\boldsymbol \beta^{* T}, \boldsymbol b^{* T}, \boldsymbol \psi^{* T})^T = \text{argmin}_{\boldsymbol \theta}\text{E}_{\boldsymbol\theta}[-\ell(\boldsymbol \theta)]\)
where \(\ell(\boldsymbol \theta)\) is the observed
log-likelihood across all \(K\) groups such that
\(\ell(\boldsymbol \theta) = \sum_{k=1}^K \ell_k(\boldsymbol \theta)\), where
\(\ell_k(\boldsymbol \theta) = (1/n_k) \log \int f(\boldsymbol d_k | \boldsymbol X_k, \boldsymbol \alpha_k; \boldsymbol \theta) \phi(\boldsymbol \alpha_k) d \boldsymbol \alpha_k\).

Our primary goal is to select the
true nonzero fixed and random effects, i.e. identify the set \begin{equation*}
  S = S_1 \cup S_2 = \{j: \beta_l^* \ne 0 \} \cup \{t: ||\boldsymbol b_t^*||_2 \ne 0\},
\end{equation*} where \(S_1\) and \(S_2\) represent the true fixed and random effects, respectively. When \(\boldsymbol b_t = \boldsymbol 0\),
this indicates that the effect of covariate \(t\) is fixed across the \(K\)
groups (i.e. the corresponding $t$-th row and column of $\boldsymbol \Sigma$ is set to $\boldsymbol 0$).

Our objective is to solve the penalized likelihood problem of (\ref{eqn:chp5_penlik}): \begin{equation}
  \widehat{\boldsymbol\theta} = \text{argmin}_{\boldsymbol\theta} - \ell (\boldsymbol\theta) + \lambda_0 \sum_{l=1}^{p} \rho_0 \left (\beta_l \right ) + \lambda_1 \sum_{t=1}^{q} \rho_1 \left (||\boldsymbol b_t||_2 \right ),
  \label{eqn:chp5_penlik}
\end{equation} where \(\ell(\boldsymbol \theta)\) is the observed log-likelihood for all \(K\) groups,
\(\rho_0(t)\) and \(\rho_1(t)\) are general folded-concave penalty
functions, and \(\lambda_0\) and \(\lambda_1\) are positive tuning
parameters. The penalty functions applied to the fixed effects, represented by \(\rho_0(t)\), could include the \(L_1\) penalty (LASSO), the SCAD penalty (Smoothly Clipped Absolute Deviation),
and the MCP penalty (Minimax Concave Penalty).\citep{glmnet2010, ncvreg2011} The penalty functions applied to the random effects, represented by \(\rho_1(t)\), could include the group LASSO, the group MCP, or the group SCAD penalties presented by Breheny and Huang \cite{grpreg2015}
since we treat the elements of $\boldsymbol b_t$ as a group and penalize them in a groupwise manner. As a result, these groups of $\boldsymbol b_t$ are estimated to be either all zero or all nonzero, which means that we
select covariates to have random effects
(\(\boldsymbol{\widehat b}_t \ne \boldsymbol 0\)) or fixed effects
(\(\boldsymbol{\widehat b}_t = \boldsymbol 0\)) across the \(K\)
groups.


Before moving on to the specifics of the algorithm used to perform this variable selection procedure and estimate the fixed and random effect parameters of interest, we want to provide some further clarifications about the values of $p$, $q$, and $r$ discussed in both this section and remaining sections. The values of $p$ and $q$ refer to the full set of fixed and random effect predictors, respectively, that are considered in the model and would be input into the \textbf{phmmPen\_FA} variable selection procedure. 
Let us represent the true number of non-zero fixed and random effects of in the model as $p^{*} \leq p$ and $q^{*} \leq q$, respectively. When we claim that $r \ll q$ (i.e. $r$ is much less than $q$), this refers to the relative size of the number of latent factors used in the model versus the total number of random effects considered in the model. We do not necessarily assume $r \ll q^{*}$ holds for the true number of random effects in the model, although we do assume $r < q^{*}$.

\subsection{MCECM algorithm}
\label{sec:proj3_mcecm}

We solve (\ref{eqn:chp5_penlik}) for some specific $(\lambda_0,\lambda_1)$ using a Monte Carlo Expectation
Conditional Minimization (MCECM) algorithm.\citep{garcia2010} 

Our objective within the \(s^{th}\) iteration of the MCECM algorithm is to evaluate the expectation of (E-step) and minimize (M-step) the penalized Q-function defined in (\ref{eqn:chp5_Qfun}): \begin{align}
  \begin{aligned}
    Q_\lambda(\boldsymbol\theta | \boldsymbol \theta^{(s)}) & = \sum_{k=1}^K E \left \{ -\log(f(\boldsymbol d_k, \boldsymbol X_k, \boldsymbol\alpha_k;\boldsymbol\theta | \boldsymbol D_o; \boldsymbol\theta^{(s)})) \right \} + \lambda_0 \sum_{j=1}^{p} \rho_0 \left (\beta_j \right ) + \lambda_1 \sum_{t=1}^{q} \rho_1 \left (||\boldsymbol b_t||_2 \right ) \\
    & = Q_1(\boldsymbol\theta | \boldsymbol\theta^{(s)}) + Q_2(\boldsymbol\theta^{(s)}) + \lambda_0 \sum_{j=1}^{p} \rho_0 \left (\beta_j \right ) + \lambda_1 \sum_{t=1}^{q} \rho_1 \left (||\boldsymbol b_t||_2 \right ),
    \label{eqn:chp5_Qfun}
  \end{aligned}
\end{align} where
\((\boldsymbol d_k, \boldsymbol X_k, \boldsymbol \alpha_k)\) gives the
complete data for group $k$,
\(\boldsymbol D_{k,o} = (\boldsymbol d_k, \boldsymbol X_k)\) gives the observed data
for group $k$, \(\boldsymbol D_o\) represents the entirety of the observed
data, and \begin{equation}
  Q_1(\boldsymbol\theta | \boldsymbol\theta^{(s)}) = - \sum_{k=1}^K \int \log [f(\boldsymbol d_k | \boldsymbol X_k, \boldsymbol\alpha_k; \boldsymbol\theta)] \phi(\boldsymbol\alpha_k | \boldsymbol D_{k,o}; \boldsymbol\theta^{(s)}) d \boldsymbol\alpha_k,
  \label{eqn:chp5_Q1}
\end{equation} \begin{equation}
   Q_2(\boldsymbol\theta^{(s)}) = - \sum_{k=1}^K \int \log [\phi(\boldsymbol \alpha_k)] \phi(\boldsymbol\alpha_k | \boldsymbol D_{k,o}; \boldsymbol\theta^{(s)}) d \boldsymbol\alpha_k
  \label{eqn:chp5_Q2}
\end{equation}
where $f(\boldsymbol d_k | \boldsymbol X_k, \boldsymbol\alpha_k; \boldsymbol\theta)$ was defined in (\ref{eqn:chp5_PE_likelihood}) and $\phi(\boldsymbol\alpha_k | \boldsymbol D_{k,o}; \boldsymbol\theta^{(s)})$ represents the posterior distribution of the latent factors $\boldsymbol \alpha_k$.

Our goal in the E-step of the algorithm is to approximate the $r$-dimensional integral expressed in (\ref{eqn:chp5_Q1}). We first specify $J$ time intervals (see Section \ref{sec:proj3_sims} for a discussion on choosing a value of $J$) defined so that there are an approximately equal number of events within each time interval.\citep{allison2010survival} If a subject survived at least part-way through $j^*$ intervals (i.e. $t^*_{kij} > 0$ for $j=1,...,j^* \leq J$), the long-form dataset contains $j^*$ observations for that subject. For subject $i$ in group $k$ that survived at least part-way through $j^*$ time intervals, we define $d_{kij} = I(\tau_{j-1} \leq y_{ki} < \tau_j, \delta_{ki} = 1)$ for $j=1,...,j^* \leq J$, the subject's $\boldsymbol x_{ki}$ and $\boldsymbol z_{ki}$ covariates are repeated for all $j^*$ observations, the $\log(t^*_{kij})$ offset term is calculated for each interval, and additional reference coded indicator values $\boldsymbol v_{kij} = (v_{kij,1},...,v_{kij,J})^T$ for the time interval $j=1,...,j^*$ are specified. The first element $\boldsymbol v_{kij}$ (element $v_{kij,1}$) is always 1, encoding a fixed effect intercept which represents time interval 1. For time interval $j>1$, the $j$-th element $\boldsymbol v_{kij}$ (element $v_{kij,j}$) is also 1. All other elements of $\boldsymbol v_{kij}$ are set to 0. 

Instead of estimating $\boldsymbol \psi$ directly, we reformulate this quantity as $\boldsymbol{\tilde \psi}$, where $\psi_1 = \tilde \psi_1$ and $\psi_j = \tilde \psi_1 + \tilde \psi_j$ for $j\geq 2$. In this formulation, $\tilde \psi_1$ estimates the log of the baseline hazard for time interval $[\tau_0,
\tau_1)$, and $\tilde \psi_1 + \tilde \psi_j$ estimates the log of the baseline hazard for time interval $[\tau_{j-1},
\tau_j)$ for $j=2,...,J$. By estimating the log baseline hazard parameters in this way, we are including a fixed effect intercept in our model.  By including a fixed effect intercept, we ensure that the full $\boldsymbol z_{ki}$ vector, which includes a random intercept, is a subset of the subject's fixed effects. 

We can re-write the log-linear model of (\ref{eqn:chp5_pois_model}) as
\begin{equation}
    \log \mu_{kij} = \log t^*_{kij} + \boldsymbol v_{kij}^T \boldsymbol{\tilde \psi} + \boldsymbol x_{ki}^T \boldsymbol \beta + \boldsymbol z_{ki}^T \boldsymbol B \boldsymbol \alpha_{k}.
    \label{eqn:chp5_pois_model_star}
\end{equation}


\subsubsection{Monte-Carlo E-step}
\label{sec:estep}
The integrals in the Q-function do not have closed forms. We approximate these integrals using a Markov
Chain Monte Carlo (MCMC) sample of size M from the posterior density
\(\phi(\boldsymbol \alpha_k | \boldsymbol D_{k,o}; \boldsymbol \theta^{(s)})\). 
Let \(\boldsymbol \alpha_k^{(s,m)}\) be the
\(m^{th}\) simulated $r$-dimensional vector from the posterior of the latent common factors, \(m = 1,...,M\), at the \(s^{th}\) iteration
of the algorithm for group \(k\). The integral in (\ref{eqn:chp5_Q1}) can be approximated as \begin{align}
\begin{aligned}
    Q_1(\boldsymbol\theta | \boldsymbol\theta^{(s)}) &\approx - \frac{1}{M} \sum_{m=1}^M \sum_{k=1}^K \log f(\boldsymbol d_k | \boldsymbol X_k, \boldsymbol\alpha_k^{(s,m)}; \boldsymbol\theta) \\
  &=  - \frac{1}{M} \sum_{m=1}^M \sum_{k=1}^K \sum_{i=1}^{n_k} \sum_{j=1}^J I(t^*_{kij} > 0) \left [ d_{kij} \log \mu_{kij}^{(s,m)} - \mu_{kij} ^{(s,m)}\right ],
\end{aligned}  
\end{align}
where $\log \mu_{kij}^{(s,m)} = \log t^*_{kij} + \boldsymbol v_{kij}^T \boldsymbol{\tilde \psi} + \boldsymbol x_{ki}^T \boldsymbol \beta + \boldsymbol z_{ki}^T \boldsymbol B \boldsymbol \alpha_{k}^{(s,m)}$. We use the No-U-Turn Sampler Hamiltonian Monte Carlo sampling procedure (NUTS HMC) from the Stan software \citep{stan2017, hoffman2014NUTS} to sample the latent factors $\boldsymbol{\alpha}_k^{(s,m)}$; we use this sampling procedure (implemented using the \textbf{rstan} R package \citep{stan2020}) because this helps improve the overall speed and efficiency of our E-step compared to other appropriate sampling method options, thereby helping to improve the speed of the overall MCECM algorithm. While we used the NUTS HMC sampling procedure for all of the analyses described in this paper, our software (see the Supporing Information section) allows for other MCMC sampling procedures including the Metropolis-within-Gibbs with
an adaptive random walk sampler \citep{adaptMCMC2009} and the
Metropolis-within-Gibbs with an independence sampler.\citep{compstats2012}

\subsubsection{M-step}
\label{sec:proj3_mstep}
In the M-step of the algorithm, we aim to minimize \begin{equation}
  Q_{1,\lambda}(\boldsymbol\theta | \boldsymbol\theta^{(s)}) = Q_1(\boldsymbol\theta | \boldsymbol\theta^{(s)}) + \lambda_0 \sum_{l=1}^{p} \rho_0 \left (\beta_l \right ) + \lambda_1 \sum_{t=1}^{q} \rho_1 \left (||\boldsymbol b_t||_2 \right )
  \label{eqn:chp5_Mstep}
\end{equation} with respect to
\(\boldsymbol \theta = (\boldsymbol \beta^T, \boldsymbol b^T, \boldsymbol{\tilde \psi}^T)^T\). We do this by using a Majorization-Minimization algorithm with penalties applied to the fixed effects $\boldsymbol \beta$ and the rows of $\boldsymbol B$. The step size of the Majorization-Minimization algorithm is estimated using a proximal gradient line search algorithm.\citep{parikh2014proximal}

Let $s$ represent the iteration of the MCECM algorithm, and let $g$ represent the iteration within a particular M-step of the MCECM algorithm such that the coefficients for the $g$-th M-step update within the $s$-th MCECM iteration is $\boldsymbol{\theta}^{(s,g)}$.

\textbf{Initialization}: We initialize the parameters $\boldsymbol{\theta}^{(s,0)}$ using $\boldsymbol{\theta}^{(s-1)}$. The step size for the Majorization-Minimization algorithm, $c^{(s,0)}$, is initialized using $c^{(s-1)}$ (value of 1.0 if $s=1$). Details on the initialization of $\boldsymbol{\theta}^{(0)}$ is given in Section \ref{sec:mcecm_alg}.

\textbf{Coefficient Updates}: Conditional on $\boldsymbol \beta^{(s,g-1)}$ and $\boldsymbol b^{(s,g-1)}$, each $\tilde \psi_j^{(s,g)}$ for $j = 1,...,J$ is given a single update using the Majorization-Minimization algorithm specified by Breheny and Huang \cite{grpreg2015} with no penalization applied.

Conditional on $\boldsymbol b^{(s,g-1)}$ and the recently updated $\boldsymbol{\tilde \psi}^{(s,g)}$, each $\beta_l^{(s,g)}$ for $l = 1,...,p$ is given a single update using the Majorization-Minimization algorithm specified by Breheny and Huang.\cite{grpreg2015}

Conditional on the recently updated $\boldsymbol \beta^{(s,g)}$  and $\boldsymbol{\tilde \psi}^{(s,g)}$, each $\boldsymbol b_t^{(s,g)}$ for $t = 1,...,q$ is updated using the Majorization-Minimization coordinate descent grouped variable selection algorithm specified by Breheny and Huang.\cite{grpreg2015}

If necessary, the step size $c^{s,g+1}$ is updated using a proximal gradient line search algorithm \citep{parikh2014proximal} and multiplied by a factor of 0.95. 

\textbf{Convergence}: The coefficient update steps are repeated until the convergence criteria specified in Supplementary Material Section 1.4 is reached or until the M-step reaches its maximum number of iterations (default 50).

\subsubsection{MCECM algorithm} 
\label{sec:mcecm_alg}
The full MCECM algorithm for estimating the parameters with a particular $(\lambda_0,\lambda_1)$ proceeds as described below. 

\textbf{Initialization}: Very briefly, we initialize $\boldsymbol{\theta}^{(0)}$ using either the coefficients from a previous model fit or from a naive model assuming no random effects if no previous model fit is available. We include additional comments on initialization later in this section; see Supplementary Material Section 1.4 for full details.

\textbf{E-step}:  For EM iteration $s$, a burn-in sample from the posterior distribution of the latent factors is run and discarded. A sample of size $M^{(s)}$ from the posterior ($\boldsymbol{\tilde \alpha}_k^{(s)} = ((\boldsymbol \alpha_k^{(s,1)})^T,...,(\boldsymbol \alpha_k^{(s,M)})^T)^T$ for $k=1,...,K$) is then drawn and retained for the M-step. 

\textbf{M-step}: Parameter estimates of $\boldsymbol{\tilde \psi}^{(s)}$, $\boldsymbol{\beta}^{(s)}$, and $\boldsymbol{b}^{(s)}$ are then updated as described in the M-step procedure given in Section \ref{sec:proj3_mstep}.

\textbf{Convergence}: The E-step and M-step are repeated until the MCECM convergence condition specified in Supplementary Material Section 1.4 is met two consecutive times (default) or until the maximum number of EM iterations is reached (25 in our simulations).


Supplementary Material Sections 1.2 and 1.3 outline the process of model selection and finding optimal tuning parameters. In brief, the algorithm runs a computationally efficient two-stage approach to pick the optimal set of tuning parameters. In the first stage of this approach, the algorithm keeps the fixed effect penalty constant at the minimum value of the fixed effects penalty sequence \(\lambda_{0,min}\), and searches over the sequence of the random effects penalties from the minimum random effect penalty \(\lambda_{1,min}\) to the maximum value \(\lambda_{1,max}\). We use model selection criteria (BIC-ICQ \citep{BICq2011}) to select the best random effect penalty, or \(\lambda_{1,opt}\). In the second stage, the algorithm keeps the random effect penalty fixed at \(\lambda_{1,opt}\) and searches over the sequence of fixed effect penalties from \(\lambda_{0,min}\) to \(\lambda_{0,max}\). The overall best model is chosen from the models in the second stage using model selection criteria. Details on the calculation of the $\boldsymbol \lambda_0$ and $\boldsymbol \lambda_1$ sequences are given in Section 1.3. 

During the development of this method, we found that certain initialization procedures helped improve variable selection and estimation results. Therefore, we briefly summarize our initialization procedure here, with full details of our initialization and convergence procedures provided in Supplementary Material Section 1.4. First, we outline how fixed and random effect coefficients are initialized in the first model (using minimum penalties \(\lambda_{0,min}\) and \(\lambda_{1,min}\)) in the sequence of models fit for the variable selection procedure. 

We initialize the fixed effects coefficients (baseline constant hazard coefficient values $\boldsymbol{ \tilde \psi}^{(0)}$ and predictor fixed effect coefficient values $\boldsymbol{\beta}^{(0)}$) by first fitting a penalized piecewise constant hazard model assuming only fixed effects and no random effects. In this model, the $\boldsymbol{ \tilde \psi}^{(0)}$ coefficients are not penalized, and we penalize the $\boldsymbol{\beta}^{(0)}$ coefficients using the minimum fixed effects penalty $\lambda_{0,min}$. The coefficient values from this minimum penalty model assuming no random effects are used as input for the first model of the overall variable selection procedure.

Based on the initialized $\boldsymbol{\beta}^{(0)}$, the predictors with non-zero initialized fixed effects are also initialized to have non-zero random effects (i.e. the corresponding rows of the $\boldsymbol{B}^{(0)}$ matrix are set to non-zero values), and predictors with zero-valued initialized fixed effects are initialized to have zero-valued random effects (i.e. the corresponding rows of the $\boldsymbol{B}^{(0)}$ matrix are set to zero). This initial screening of random effects helps improve the speed of the algorithm. See Section 1.4 for details on how the non-zero elements of $\boldsymbol{B}^{(0)}$ are initialized. 

After we fit the first model in the overall variable selection procedure using the MCECM algorithm, the fixed and random effect coefficients in consecutive models are initialized using the values from the previous model fit. We found that progressing through penalty values from the minimum penalty to the maximum penalty (as discussed above and in Sections 1.2 and 1.3 of the Supplementary Material) significantly improved initialization of subsequent models, as opposed to proceeding from the maximum penalty to the minimum penalty as some other fixed-effects only penalization methods do.\citep{glmnet2010,ncvreg2011}

\subsection{Estimation of the number of latent factors}
\label{sec:proj3_lit_r_est}

Performing our proposed \textbf{phmmPen\_FA} method requires specifying the number of latent factors $r$. Since $r$ is typically unknown \emph{a priori}, this value needs to be estimated. Here, we use the Growth Ratio (GR) procedure.\citep{ahn2013eigenvalue}

The GR method for our application requires a $q \times K$ matrix of observed random effects. Since these random effects cannot be directly observed, we instead calculate pseudo random effects by first fitting a penalized piecewise constant survival model with a small penalty to each group individually using the random effect covariates of interest as the predictors in the model. We then take these group-specific estimates and center them so that all features have a mean of 0. Let these $q$-dimensional group-specific estimates be denoted as $\boldsymbol{\hat \gamma_k}$ for each group $k=1,...,K$. We then define $\boldsymbol G = (\boldsymbol{\hat \gamma_1},...,\boldsymbol{\hat \gamma_K})$ as the final $q \times K$ matrix of pseudo random effects. 

Let $\xi_j(A)$ be the $j$-th largest eigenvalue of the positive semidefinite matrix $A$, and let $\tilde \mu_{qK,j} \equiv \xi_j (\boldsymbol G \boldsymbol G^T / (qK)) = \xi_j (\boldsymbol G^T \boldsymbol G / (qK))$. To find the GR estimator, we first order the eigenvalues of $\boldsymbol G \boldsymbol G^T / (qK)$ from largest to smallest. 
Then, we calculate the following ratios:
\begin{equation}
    GR(j) \equiv \frac{\log [V(j-1) / V(j)]}{\log [V(j)/V(j+1)]} = \frac{\log (1 + \tilde \mu_{qK,j}^*)}{\log (1 + \tilde \mu_{qK,j+1}^*)}, \hspace{10pt} j = 1,2,...,U
\end{equation}
where $V(j) = \sum_{l=j+1}^{\min(q,K)} \tilde \mu_{qK,l}$, $\tilde \mu_{qK,j}^* = \tilde \mu_{qK,j} / V(j)$, and $U$ is a pre-defined constant. Then, we estimate $r$ by
\begin{equation}
    \widehat{r}_{GR} = \text{max}_{1 \leq j \leq U} GR(j)
\end{equation}


\section{Simulations}
\label{sec:proj3_sims}

In this section, we examine how well the \textbf{phmmPen\_FA} algorithm performs variable selection on the fixed and random effects covariates for piecewise constant hazard mixed effects models under several different conditions. In all of these simulations, 
we use the MCP penalty (MCP penalty for the fixed effects, group MCP penalty for the rows of the $\boldsymbol B$ matrix) and the BIC-ICQ \citep{BICq2011} model selection criterion with the abbreviated two-stage grid search as described in the Section \ref{sec:proj3_mcecm} (see full details in Supplementary Material Section 1.2). In order to determine the robustness of our variable selection procedure based on the assumed value of $r$, we fit models in one of two ways: we estimated the number of common factors $r$ using the Growth Ratio estimation procedure, or we use the true value of $r$.

In all of our simulations, we specified $J=8$ time intervals for the piecewise constant hazard mixed effect survival model; we chose $J=8$ because this is the default number of time intervals used within the piecewise constant hazard procedure implemented in the SAS \textbf{proc phreg} command.\citep{SAS_phreg} We created the intervals such that that there were an approximately equal number of events within each time interval, as suggested by Allison.\cite{allison2010survival}
There are several alternative options for choosing the number of time intervals to include in a piecewise constant hazard mixed effect model. 
Castillo and van der Pas \cite{castillo2021multiscale} proposed using $J=\sqrt{n/\log n}$ as a data-driven estimate for the number of intervals. Our review of the literature suggests that it is common to choose between 5 and 10 time intervals for piecewise constant hazard models when they are fitting real data.\citep{austin2017tutorial, SAS_phreg, li2012piecewise, qiou1999multivariate,friedman1982piecewise} One could try several values of $J$ and compare the models using appropriate Bayesian model selection criteria, including the DIC,\citep{spiegelhalter2002bayesian,cancho2011flexible,ibrahim2012bayesian}
the $L$ measure,\citep{ibrahim2001bayesian,ibrahim2001criterion} or the BIC-ICQ.\citep{BICq2011} 
In general, it is suggested to use a moderate number of time intervals because using too many time intervals can result in unstable estimates, but too few time intervals can lead to an inadequate model fit.\citep{qiou1999multivariate}
In the Supplementary Materials Section 3.5, we repeat some simulations assuming a range between 5 and 10 time intervals in our \textbf{phmmPen\_FA} procedure. The variable selection performance was very similar across this range of time intervals, see the Supplementary Material for full details. Our software in the package \textbf{phmmPen\_FA} recommends choosing between 5 and 10 time intervals and specifies a default of 8 time intervals.


\subsection{Variable selection in survival data with 100 predictors}
\label{sec:proj3_sim100}

We examine the performance and scalability of the \textbf{phmmPen\_FA} algorithm when performing variable selection in high dimensions of $p=100$ total predictors. We simulated survival data from a piecewise constant hazard mixed effect model with $p$ predictors. Of \(p\) total predictors, we assume that the first 5 predictors have truly non-zero fixed and random effects (i.e. $p^* = 5$ and $q^*=p^*+1$, where the additional $1$ comes from including a random intercept to account for group-specific variations in the baseline hazard), and the other \(p-5\) predictors have zero-valued fixed and random effects. We specified a full model for the algorithm such that the random effect predictors equalled the fixed effect predictors (e.g. in the full model $q=p+1$), and our aim was to select the set of true predictors and random effects.

To simulate the data, we set the total sample size to $N=1000$ and considered the number of groups $K$ to be either 5 or 10, with an equal number of subjects per group. We set up the random effects covariance matrix by specifying a $\boldsymbol B$ matrix with dimensions $(p+1)\times r$, where $p+1$ represents the $p$ predictors specified in the $\boldsymbol X$ matrix plus the random intercept, and the number of latent common factors $r$ was set to three. Six of these $p+1$ rows---corresponding to the true 5 predictors plus the random intercept---had non-zero elements, while the remaining $p-5$ rows were set to zero. For each value of $r$, we considered a $\boldsymbol B$ matrix that produced $\boldsymbol \Sigma = \boldsymbol B \boldsymbol B^T$ with either small or moderate variances and eigenvalues; 
see Section 1.1 of the Supplementary Material for further details. These two cases are referred to as the `small' or `moderate' $\boldsymbol B$ matrices in the simulation results presented in this section.
We generate both moderate and strong predictor effects, where all 5 of the true fixed effects have coefficient values of 0.5 or 1.0, respectively. Each condition was evaluated using 100 total simulated datasets.

In order to sample event times $\boldsymbol T = (\boldsymbol T_1^T, ..., \boldsymbol T_k^T)^T$ where $\boldsymbol T_k = (T_{k1},...,T_{kn_k})^T$, we defined five half-year time intervals as $\{[0,0.5),[0.5,1.0),[1.0,1.5),[1.5,2.0),[2.0,\infty)\}$. The corresponding log baseline hazard values for these intervals were $\boldsymbol \psi_j^* = (-1.5,1.0,2.7,3.7,6.8)$. 

For group \(k\), we generated the event times \(T_{ki}\), \(i=1,...,n_k\), using the following procedure: We first simulated values from the exponential distribution $e_{kij} \sim Exp(R_{kij})$ starting with $j=1$, where the exponential rate $R_{kij} = \exp(\psi_j + \boldsymbol x_{ki}^T \boldsymbol \beta + \boldsymbol z_{ki}^T \boldsymbol \gamma_k)$, where \(\boldsymbol \gamma_k \sim N_{6}(0, \boldsymbol B \boldsymbol B^T)\). If the inequality $\tau_j < \tau_{j-1} + e_{kij}$ was true, then we simulated $e_{kij}$ using the $j+1$ interval parameters until either the inequality $\tau_j >= \tau_{j-1} + e_{kij}$ held for a particular $j^*$ or the last time interval $J$ was reached. We then defined $T_{ki} = e_{kij^*} + \tau_{j^*-1}$. 

Censoring times $\boldsymbol C = (\boldsymbol C_1^T, ..., \boldsymbol C_k^T)^T$ where $\boldsymbol C_k = (C_{k1},...,C_{kn_k})^T$ were simulated from the uniform distribution $C_{ki} \sim Unif(0,5)$, which assumes an end to follow-up after 5 years. The aforementioned event time simulation in combination with this censoring time simulation resulted in censoring rates that fell within 11\% to 26\% for all simulation conditions; average censoring rates across the 100 simulation replicates for the conditions mentioned ranged from 16\% to 19\%. The average median follow-up time was approximately 0.60 years for the various conditions.

For individual \(i\) in group \(k\), the vector of predictors for the fixed effects is given as \(\boldsymbol x_{ki} = (x_{ki,1},...,x_{ki,p})^T\), which does not inlcude an intercept, and we define the random effects \(\boldsymbol z_{ki} = (1,\boldsymbol x_{ki})\), where \(x_{ki,l} \sim N(0,1)\) for \(l=1,...,p\), and each $\boldsymbol x_l$ was standardized as described in Section \ref{sec:proj3_models}. We include a random intercept in the random effects predictors $\boldsymbol z_{ki}$ to allow for the baseline hazard to vary across groups. 

We prepared the data to be fit with a piecewise constant hazard survival model by calculating eight time intervals---specified such that there were an approximately equal number of events within each time interval---and then creating the long-form dataset specified in Section \ref{sec:proj3_mcecm} using the \texttt{survival::survSplit()} function from the \textbf{survival} R package.\citep{survival_package,survival_pkg_book} 


The results for these simulations are presented in Tables \ref{tab:p100_pos} and \ref{tab:p100_r}. Table \ref{tab:p100_pos} provides the average true and false positive percentages for both the fixed and random effects variable selection, the median time in hours to complete the variable selection procedure, the average of the mean absolute deviation between the fixed effects coefficient estimates and the true fixed effects coefficients across all simulation replicates, and the average of the Frobenius norm of the difference between the estimated random effect covariance matrix $\boldsymbol{\hat \Sigma} = \boldsymbol{\hat B \hat B^T}$ and the true covariance matrix $\boldsymbol{\Sigma} = \boldsymbol{B B^T}$ (the Frobenius norm was standardized by the number of random effects selected in the best model). The true positive percentages express the average percent of the true predictors selected in the best models across simulation replicates, and the false positive percentages express the average percent of false predictors selected in the best models. Table \ref{tab:p100_r} gives the Growth Ratio estimation procedure results, including the average estimate of $r$ and the proportion of times that the Growth Ratio estimate of $r$ was underestimated, correct, or overestimated. All simulations were completed on a high performance computing cluster with CPU Intel processors between 2.3Ghz and 2.5GHz. 

We see from Table \ref{tab:p100_pos} that the \textbf{phmmPen\_FA} method is able to accurately select both the fixed and random effects within the piecewise constant hazard mixed effects model across a variety of conditions. The true positive rates of the \textbf{phmmPen\_FA} method are generally above 90\% for both fixed and random effects; the fixed effects true positives increase when the true predictor effects are larger, and the random effects true positives increase when the number of groups in the data increase. The false positive rates are less than 6.5\% for fixed effects and less than 3.9\% for the random effects across all conditions. 

We can see from Table \ref{tab:p100_r} that the Growth Ratio estimation procedure generally underestimates the number of latent factors $r$ for the simulated data set-ups used in this section. We expect that this is a result of a combination of reasons, including relatively low numbers of groups $K$ in the data and $\boldsymbol B$ matrices that created $\boldsymbol \Sigma$ matrices with relatively low eigenvalues. This is supported by the results that show an improvement in the accuracy of the estimation as the number of groups and the relative size of the $\boldsymbol B$ matrix increases. Additionally, the Growth Ratio utilizes group-specific penalized piecewise constant hazard coefficient estimates, and these estimates might be sensitive to the fact that for some simulated datasets, not all groups had a sufficient number of events within each time interval to get reasonable $\boldsymbol{\tilde \psi}$ estimates, possibly leading to less than stable pseudo random effect estimates. See a further discussion on this topic in Sections 4.1 and 3.5 of the Supplementary Material.

Even though the Growth Ratio procedure consistently underestimated $r$, this did not strongly impact the variable selection results nor the bias of the fixed effects estimates selected in the best models. When the algorithm used the Growth Ratio estimate of $r$ instead of the true estimate of $r$, the true and false positive rates remained very consistent, with only slight decreases in true positive rates for the fixed and random effects when the Growth Ratio procedure is used. The largest impact that underestimating $r$ had on the bias of the fixed effects estimates was when $K=10$ and $\beta=1.0$.

Our observation in these simulations that underestimating the number of latent factors $r$ does not harm our method’s performance compared with using the true $r$ can be explained by the eigenvalues of the random effects covariance matrix $\boldsymbol{\Sigma}$. Our assumption that $\boldsymbol{\Sigma}$ can be represented by the low-rank (rank $r$) matrix of $\boldsymbol{B B}^T$  (i.e. can be represented with $r$ latent factors) means that $\boldsymbol{\Sigma}$ has $r$ non-zero eigenvalues. The sizes of these eigenvalues provide an indication of how important each latent factor is for representing $\boldsymbol{\Sigma}$. Suppose we order these $r$ eigenvalues from largest to smallest. If the first $r-1$ eigenvalues are relatively large and the $r$-th eigenvalue is relatively small, this means that most of the variation in $\boldsymbol{\Sigma}$ can be represented by the first $r-1$ latent factors (i.e. the first $r-1$ columns of the $\boldsymbol{B}$ factor loadings matrix).

In our simulations, the moderate $\boldsymbol{B}$ results in eigenvalues (3.38, 3.38, 2.25), and the small $\boldsymbol{B}$ results in eigenvalues (1.5, 1.5, 1.0) (see details of the $\boldsymbol{B}$ matrices in Supplementary Material Section 1.1). Since the third eigenvalue is smaller than the first two, this indicates that the first two latent factors are the most important, and the third latent factor is of lesser importance. Therefore, if we underestimate r as 2 instead of 3 in our simulations, then we are perhaps not losing a large amount of information when we try and estimate the random effect covariance matrix $\boldsymbol{\Sigma}$. 

\subsection{Variable selection in survival data with 500 predictors}
\label{sec:proj3_sim500}
In order to further illustrate the scalability of our method, we applied our method to survival simulations with $p=500$ covariates. We simulated the event and censoring times from a piecewise constant hazard mixed effects model much like the procedure described in Section \ref{sec:proj3_sim100}, except the total number of predictors used in the analyses was $p=500$ instead of $p=100$. All simulations assumed the true number of latent factors $r$ was 3 and the Growth Ratio method was used to estimate $r$. Just as in the $p=100$ survival simulations, we specified a full model for the algorithm such the random effect predictors equalled the fixed effect predictors (~e.g. $q=p+1$), and our aim was to select the set of true predictors and random effects. The variable selection results to these simulations are given in Table \ref{tab:p500_pos}. The median times needed to complete these simulations took between 11.9 and 21.3 hours.

When we compare specific data conditions (i.e. specific combinations of the size of the fixed and random effects and the number of groups) between Tables \ref{tab:p100_pos} and \ref{tab:p500_pos}, we see that increasing the number of total predictors input into the \textbf{phmmPen\_FA} procedure from $p=100$ to $500$ tended to decrease the true positive rates for both the fixed and random effects and increase the total time needed to complete the variable selection procedure. Similar to the $p=100$ simulations, the Growth Ratio procedure continued to underestimate the number of latent factors in the underlying model.

\subsection{Supplemental simulations}
\label{sec:sim_supp_details}

The Supplementary Materials Section 3 provides results from additional simulations. Supplementary Material Section 3.1 compares our \textbf{phmmPen\_FA} method to a naive fixed-effects only variable selection approach (i.e. no random effects incorporated into the survival model), implemented using the \textbf{ncvreg} R package.\citep{ncvreg2011} In terms of selecting the fixed effects predictors, the \textbf{phmmPen\_FA} method outperforms the naive fixed-effects only method of \textbf{ncvreg} when the true fixed effects coefficients are more moderate or when the true random effect coefficients are larger; \textbf{ncvreg} performs comparably to \textbf{phmmPen\_FA} in situations when the true fixed effects coefficients are larger and the random effects coefficients are smaller. As expected, the \textbf{ncvreg} package has the advantage of performing variable selection much faster since no random effects need to be selected. 

Supplementary Material Section 3.2 illustrates how the \textbf{phmmPen\_FA} method performs when the mixed effects survival data is simulated using the Weibull distribution instead of the piecewise constant hazard distribution. The results indicate that our method performs comparably for either data generation mechanism, and our method even performs slightly better on the Weibull-generated data according to some metrics. 

Supplementary Material Section 3.3 examines how the \textbf{phmmPen\_FA} method performs when the true underlying data has a different number of true fixed effects predictors and true random effects predictors. The results show that our method can also accurately select the fixed and random effects predictors in these scenarios. 

Supplementary Material Section 3.4 further explores how the \textbf{phmmPen\_FA} method performs when the number of latent factors $r$ is purposefully underestimated. In general, decreasing $r$ helps decrease the time needed to complete the \textbf{phmmPen\_FA} procedure. As the value of $r$ used in the algorithm decreases and deviates further from the true value of $r$, the true positives of the fixed and random effects generally decreases; the false positives of the random effects generally increases; and the bias of the random effects covariance matrix generally increases. However, the differences between variable selection and bias performances across values of $r$ continue to be fairly minor. 

Supplementary Materials Section 3.5 investigates how the \textbf{phmmPen\_FA} method performs when different numbers of time intervals $J$ are assumed (values of $J$ from 5 to 10 were considered). In general, there was very little difference in the variable selection performance between different values of $J$. In the data scenarios considered in these simulations, smaller $J$ values improved the accuracy of the Growth Ratio $r$ estimate. Smaller $J$ also decreased the median time needed to complete the procedure.

\section{Case study: Pancreatic Ductal Adenocarcinoma}
\label{sec:proj3_PDAC}

Patients diagnosed with Pancreatic Ductal Adenocarcinoma (PDAC) generally face a very poor prognosis, where the 5-year survival rate is 6\%.\citep{khorana2016potentially} Consequently, it is of clinical interest to robustly identify gene signatures that are associated with overall survival to better predict patient prognosis in the clinic.

Selecting gene signatures for the prediction of clinical outcomes, including survival outcomes, can often be inconsistent across biomedical studies, where gene signatures identified in one study may have little or no overlap with ones identified in other studies.\citep{waldron2014} Consequences of this lack of replicability in gene signature selection include variable accuracy in predicting clinical outcomes in new studies using these models \citep{sotiriou2007,waldron2014} and contradictory effect estimates relating genes to the outcome.\citep{swisher2012molecular} This lack of replicability across studies can come from small sample size \citep{sotiriou2007} and differences in data pre-processing steps,\citep{lusa2007,paquet2015absolute} among other sources.

In order to improve replicability in the prediction of survival in PDAC, we combine PDAC gene expression data from seven different studies.\citep{aguirre2018real,cao2021proteogenomic,dijk2020unsupervised,moffitt2015,bailey2016genomic,puleo2018stratification,raphael2017integrated} The studies used in these analyses are summarized in Supplementary Material Table 1 (within Supplementary Material Section 2.1). The seven combined studies resulted in a sample size of 879 subjects with 539 events. In order to account and adjust for between-study heterogeneity, we apply our new method \textbf{phmmPen\_FA} to fit a penalized piecewise constant hazard mixed effects model to our data to select predictors with study-replicable effects, where we assume that predictor effects may vary between studies. 

Moffitt et al. \cite{moffitt2015} identified a 500-member gene list relevant to classifying two PDAC tumor subtypes they identified---basal and classical---which were prognostic of survival. Therefore, we decided to limit our initial interest to these 500 genes.  
Of these 500 genes, 420 of these genes were common among all of the datasets. We removed 20\% of the genes with the lowest gene expression based on their average rank, leaving 336 genes.


We integrated gene expression data from multiple studies by first using the data integration rank transformation technique as specified by Rashid et al.,\cite{rashid2020} allowing us to sidestep complex questions regarding how to cross-normalize data. 
This integration technique creates top scoring pairs (TSPs). To illustrate the interpretation of TSPs, let $g_{ki,A}$ and $g_{ki,B}$ be the raw expression of genes $A$ and $B$ in subject $i$ of group $k$. 
For each gene pair ($g_{ki,A}$, $g_{ki,B}$), the TSP is an indicator $I(g_{ki,A} > g_{ki,B})$ which specifies which of the two genes has higher expression in the subject. We denote a TSP predictor as ``GeneA\_GeneB''. In the dataset, we use 168 TSP predictors. The Supplementary Material Section 2.1 provides additional details on the data processing and selection of the TSPs used in the analysis.

Due to the presence of several pairwise Spearman correlation values greater than 0.5 between the TSP covariates used within the analyses, we used the Elastic Net penalization procedure \citep{glmnet2010} to balance between ridge regression and the MCP penalty (regular MCP penalty \citep{ncvreg2011} for the fixed effects, and grouped MCP penalty \citep{grpreg2015} for the rows of the \textbf{B} matrix). We let $\pi$ represent the balance between ridge regression and the MCP penalty, where $\pi=0$ represents ridge regression, $\pi=1$ represents the MCP penalty, and $\pi\in(0,1)$ represents a combination of the two. 

We used the \textbf{phmmPen\_FA} procedure to fit this PDAC survival data with a penalized piecewise constant hazard mixed effect survival model. Like in the simulations presented in Section \ref{sec:proj3_sims}, we assumed $J=8$ time intervals, with the interval boundaries selected such that an approximately equal number of events within each time interval. We considered values of $\pi=\{0.7,0.8,0.9,1.0\}$, where the same value of $\pi$ was used for both the fixed and random effects penalization, and $r$ evaluated using the Growth Ratio procedure (evaluated as 2 for all cases) and $r$ manually set to a larger value of 3 since our simulations indicated that the GR estimate of $r$ may be underestimated. The sequence of $\lambda$ penalties used in the variable selection procedure is described in the Supplementary Material Section 1.3. The best model within each $\pi$ and $r$ combination was selected using the BIC-ICQ model selection criteria.

To evaluate the performance of \textbf{phmmPen\_FA} under the various $\pi$ and $r$ combinations described above, we utilized cross validation to estimate the C-index for each set-up. We randomly selected 80\% of the observations from each study to be the training dataset (with random selection stratified by events and censored observations), and the remaining 20\% of the observations were set as the test dataset. The C-index was calculated using the \textbf{intsurv::cIndex} function,\citep{intsurv-package} and the C-index risk score was calculated as the `best' model's fixed effects coefficients applied to the training dataset TSP predictors; the absence of any random effect coefficients in this risk score calculation was done in order to replicate typical real-world mixed effects scenarios, where the groups within the data used to train the models will not often equal the groups in future data to which the model will be applied. 

\begin{figure}[h!]
\centering
\includegraphics{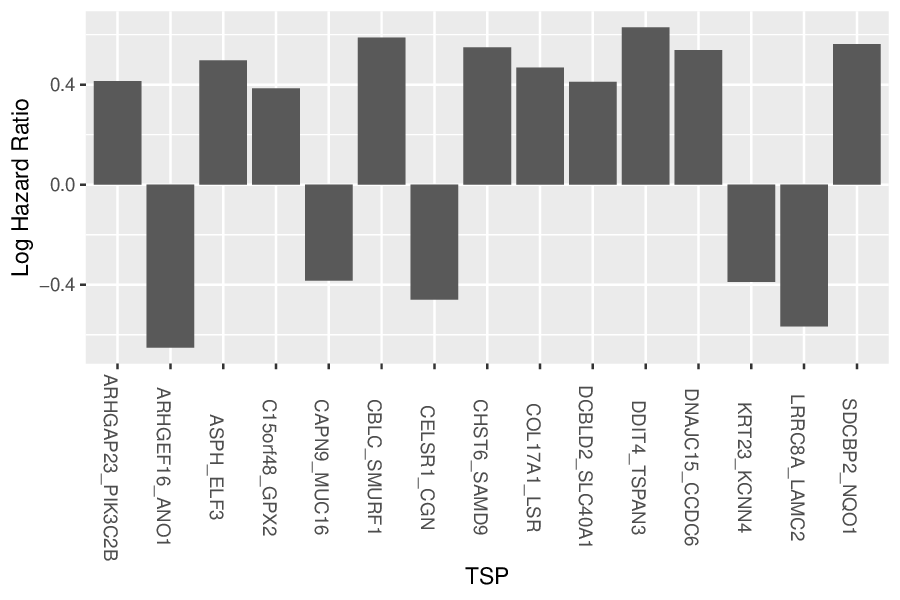}
\caption{Log hazard ratios for the TSP covariates selected during the case study analysis.}
\label{fig:PDAC_results}
\end{figure}

The combination of $\pi=0.9$ and $r=3$ produced the largest C-index value of 0.6511. This combination selected 15 of the 168 total TSPs to have non-zero fixed effects (see Figure \ref{fig:PDAC_results}) and were therefore considered important for the prediction of survival in PDAC subjects. One TSP, CBLC\_SMURF1, was selected to have a non-zero random effect (random effect variance estimated as 0.035). The time to complete the algorithm was 2.1 hours. The overall range of the C-index values across the various $\pi$ and $r$ conditions was fairly small (smallest C-index value was 0.6445 for $\pi=0.7$ and $r=2$ from the GR estimation procedure). This small range for the C-index is likely a consequence of the high proportion of censored observations within this dataset.\citep{alabdallah2024concordance}

Further details about the sensitivity analyses for this case study can be found in the Supplementary Materials Section 2.2.

\section{Discussion}
\label{sec:proj3_discuss}

We have shown through simulations and a case study of pancreatic ductal adenocarcinoma patients that we can extend the method to perform variable selection in high dimensional mixed effects models to survival data. We accomplish this by approximating proportional hazards mixed effects models using a piecewise constant hazard mixed effects model and then applying the Monte Carlo Expectation Conditional Minimization (MCECM) algorithm to simultaneously select for fixed and random effects. We incorporate the factor model decomposition of the random effects proposed in Heiling et al. \cite{heiling2024efficient} in order to scale this method to larger dimensions, e.g. hundreds of predictors.

The simulations presented in Section \ref{sec:proj3_sims} show that the \textbf{phmmPen\_FA} method can accurately select both fixed and random effects even for small or moderate effect sizes, which reflects hazard values and variations in typical survival data. By using the factor model decomposition of the random effects, this model selection procedure can be accomplished within reasonable time frames even when we consider hundreds of predictors as input for the variable selection procedure.

Our method is limited by the need to provide an estimate for the number of latent factors that model the random effects. The simulation results showed that the Growth Ratio procedure tended to underestimate this value for the simulation conditions that we considered. 
However, even when the number of latent factors was estimated incorrectly by the Growth Ratio procedure, this mis-specification had very little impact on the general variable selection performance or the fixed effects coefficient estimates. Therefore, our method is not sensitive to the estimation of the number of latent factors.

Similar to other penalization approaches to variable selection such as the \textbf{glmnet} \citep{glmnet2010} and \textbf{ncvreg} \citep{ncvreg2011} R packages, another limitation of our method is that it focuses on selecting relevant predictors but does not provide inference or measurements of uncertainty for the selected predictors. In order to get inference or uncertainty measures, one would have to take the selected model from the \textbf{phmmPen\_FA} procedure and use other software that could fit mixed effects models to provide this information, such as the \textbf{coxme} R package \citep{coxme} that fits survival mixed effects models.

In our simulations in Section \ref{sec:proj3_sims} and our case study analysis in Section \ref{sec:proj3_PDAC}, we have not compared the performance of our \textbf{phmmPen\_FA} method with other proportional hazards mixed effects (PHMM) models because we were not aware of other PHMM methods that could select both fixed and random effects and be applied to survival data with hundreds of predictors. However, we do provide a comparison between our method and a naive method that performs variable selection on fixed effects only using the \textbf{ncvreg} R package, see Supplementary Materials Section 3.1 for these results.

\bmsection*{Financial disclosure}

None reported.

\bmsection*{Conflict of interest}

The authors declare no potential conflict of interests.

\bmsection*{Supporting information}

Additional supplementary material may be found in the online version of the article at the publisher’s website, including supplementary information about the simulations presented in this paper, the \textbf{phmmPen\_FA} algorithm procedure, additional simulations not included in this paper, and the case study data processing and sensitivity analyses. Software in the form of R code for the \textbf{phmmPen\_FA} procedure is available through the \textbf{glmmPen} package in CRAN \url{https://cran.r-project.org/package=glmmPen} and the GitHub repository \url{https://github.com/hheiling/glmmPen}. The \textbf{phmmPen\_FA} procedure is implemented through the \texttt{phmmPen\_FA()} function within this \textbf{glmmPen} R package.  Code to replicate the simulation and the case study analyses and results is available through the GitHub repository \url{https://github.com/hheiling/paper_phmmPen_FA}.

\bibliography{wileyNJD-AMA}




\pagebreak
\begin{table}[h!]
\centering
\begin{tabular}{ccccccccccc}
  \hline
   $\beta$ & $K$ & $\boldsymbol B$ & $r$ Est. & 
   \multicolumn{1}{p{1.0cm}}{\centering TP \% \\ Fixef} & 
   \multicolumn{1}{p{1.0cm}}{\centering FP \% \\ Fixef} & 
   \multicolumn{1}{p{1.0cm}}{\centering TP \% \\ Ranef} & 
   \multicolumn{1}{p{1.0cm}}{\centering FP \% \\ Ranef} & $T^{med}$ &
   \multicolumn{1}{p{1.5cm}}{\centering Abs. Dev. \\ (Mean)}  &
   $||\boldsymbol D||_F$ \\
  \hline
    0.5 & 5 & Small & True & 91.80 & 2.39 & 93.00 & 0.65 & 3.81 & 0.23 & 0.31\\
     &  &  & GR  & 90.80 & 2.46 & 91.60 & 1.40 & 2.34 & 0.22 & 0.42\\
     &  & Moderate & True & 91.00 & 4.85 & 94.40 & 1.31 & 6.96 & 0.34 & 0.63\\
     &  &  & GR & 91.00 & 4.18 & 92.40 & 2.13 & 3.78 & 0.32 & 0.70\\
     & 10 & Small & True & 94.60 & 2.18 & 98.60 & 0.99 & 4.81 & 0.17 & 0.27\\
     &  &  & GR & 94.00 & 3.28 & 95.80 & 1.63 & 2.66 & 0.17 & 0.31\\
     &  & Moderate & True & 90.00 & 5.25 & 94.60 & 3.57 & 6.08 & 0.24 & 0.54 \\
     &  &  & GR & 86.20 & 6.42 & 93.40 & 3.83 & 3.02 & 0.23 & 0.61\\
     
    1.0 & 5 & Small & True & 99.20 & 1.05 & 96.00 & 0.16 & 6.01 & 0.26 & 0.31\\
     &  &  & GR & 99.00 & 1.09 & 93.20 & 0.54 & 3.50 & 0.25 & 0.39\\
     &  & Moderate & True & 97.60 & 2.92 & 95.20 & 0.65 & 8.72 & 0.36 & 0.63\\
     &  &  & GR & 95.20 & 2.75 & 94.20 & 1.22 & 3.63 & 0.34 & 0.71\\
     & 10 & Small & True & 100.00 & 1.02 & 99.60 & 0.15 & 5.14 & 0.20 & 0.27\\
     &  &  & GR & 99.80 & 1.20 & 97.00 & 0.34 & 2.97 & 0.23 & 0.30\\
     &  & Moderate & True & 98.80 & 2.39 & 99.60 & 1.01 & 7.55 & 0.27 & 0.55\\
     &  &  & GR & 98.20 & 4.22 & 98.80 & 0.66 & 4.02 & 0.33 & 0.61\\
   \hline \\
\end{tabular}
\caption{Variable selection results for the $p=100$ piecewise constant hazard mixed effects simulations, including true positive (TP) percentages for fixed and random effects, false positive (FP) percentages for fixed and random effects, the median time in hours for the algorithm to complete ($T^{med}$), and the average of the mean absolute deviation (Abs. Dev. (Mean)) between the coefficient estimates and the true $\beta$ values across all simulation replicates. Column $\boldsymbol B$ describes the general size of both the variances and eigenvalues of the resulting $\boldsymbol \Sigma = \boldsymbol{B B}^T$ random effects covariance matrix. Column `$r$ Est.' refers to the method used to specify $r$ in the algorithm: the Growth Ratio (GR) estimate or the true value of $r$. Column $||\boldsymbol D||_F$ represents the average across simulation replicates of the Frobenius norm of the difference ($\boldsymbol{D}$) between the estimated random effects covariance matrix $\boldsymbol{\hat \Sigma}$ and the true random effects covariance matrix $\boldsymbol{\Sigma}$; the Frobenius norm was standardized by the number of true random effects selected in the model.}
\label{tab:p100_pos}
\end{table}

\pagebreak

\begin{table}[h!]
\centering
\begin{tabular}{ccccccc}
  \hline
   $\beta$ & $K$ & $\boldsymbol B$ & Avg. $r$ & $r$ Underestimated \% & $r$ Correct \% & $r$ Overestimated \% \\ 
  \hline
  0.5 & 5 & Small & 2.00 & 100 & 0 & 0\\
   &  & Moderate & 2.00 & 100 & 0 & 0\\
   & 10 & Small & 2.07 & 95 & 3 & 2\\
   &  & Moderate & 2.20 & 85 & 10 & 5\\

  1.0 & 5 & Small & 2.00 & 100 & 0 & 0\\
   &  & Moderate & 2.00 & 100 & 0 & 0\\
   & 10 & Small & 2.13 & 88 & 11 & 1\\
   &  & Moderate & 2.19 & 83 & 15 & 2 \\
   \hline \\
\end{tabular}
\caption{Results of the Growth Ratio $r$ estimation procedure for $p=100$ piecewise constant hazard mixed effects simulation results, including the average estimate of $r$ across simulations and percent of times that the estimation procedure underestimated $r$, gave the true $r$, or overestimated $r$. Column $\boldsymbol B$ describes the general size of both the variances and eigenvalues of the resulting $\boldsymbol \Sigma = \boldsymbol{B B}^T$ random effects covariance matrix.}
\label{tab:p100_r}
\end{table}

\pagebreak

\begin{table}[!h]
\centering
\begin{tabular}{ccccccccccc}
  \hline
   $\beta$ & $K$ & $\boldsymbol B$ & Avg. $r$ & 
   \multicolumn{1}{p{1.0cm}}{\centering TP \% \\ Fixef} & 
   \multicolumn{1}{p{1.0cm}}{\centering FP \% \\ Fixef} & 
   \multicolumn{1}{p{1.0cm}}{\centering TP \% \\ Ranef} & 
   \multicolumn{1}{p{1.0cm}}{\centering FP \% \\ Ranef} & $T^{med}$ &
   \multicolumn{1}{p{1.5cm}}{\centering Abs. Dev. \\ (Mean)} & 
   $||\boldsymbol D||_F$\\ 
  \hline
    0.5 & 5 & Small & 2.00 & 86.40 & 1.97 & 77.40 & 0.81 & 14.19 & 0.18 & 0.39 \\
     &  & Moderate & 2.00 & 80.60 & 3.91 & 76.40 & 1.63 & 19.40 & 0.24 & 0.77\\
     & 10 & Small & 2.01 & 91.40 & 1.79 & 88.00 & 0.53 & 16.21 & 0.16 & 0.30\\
     &  & Moderate & 2.04 & 81.60 & 5.82 & 80.80 & 2.44 & 23.62 & 0.20 & 0.69\\

    1.0 & 5 & Small & 2.00 & 99.40 & 0.36 & 91.40 & 0.02 & 18.20 & 0.30 & 0.30\\
     &  & Moderate & 2.00 & 93.60 & 0.79 & 85.00 & 0.12 & 19.77 & 0.39 & 0.72\\
     & 10 & Small & 2.06 & 100.00 & 0.56 & 95.40 & 0.04 & 16.78 & 0.30 & 0.27\\ 
     &  & Moderate & 2.03 & 97.20 & 1.31 & 92.80 & 0.18 & 24.31 & 0.40 & 0.64\\  
   \hline \\
\end{tabular}
\caption{Variable selection results for the $p=500$ piecewise constant hazard mixed effects simulations, including true positive (TP) percentages for fixed and random effects, false positive (FP) percentages for fixed and random effects, the median time in hours for the algorithm to complete ($T^{med}$), and the average of the mean absolute deviation (Abs. Dev. (Mean)) between the coefficient estimates and the true $\beta$ values across all simulation replicates. Column $\boldsymbol B$ describes the general size of both the variances and eigenvalues of the resulting $\boldsymbol \Sigma = \boldsymbol{B B}^T$ random effects covariance matrix. Column `$r$ Est.' refers to the method used to specify $r$ in the algorithm: the Growth Ratio (GR) estimate or the true value of $r$. Column $||\boldsymbol D||_F$ represents the average across simulation replicates of the Frobenius norm of the difference ($\boldsymbol{D}$) between the estimated random effects covariance matrix $\boldsymbol{\hat \Sigma}$ and the true random effects covariance matrix $\boldsymbol{\Sigma}$; the Frobenius norm was standardized by the number of true random effects selected in the model.}
\label{tab:p500_pos}
\end{table}

\pagebreak

\end{document}